\documentclass[11pt]{article} 
\usepackage{a4wide}  
\RequirePackage[latin1]{inputenc}     
 
\usepackage[english]{babel} 
\usepackage{amsmath} 
\usepackage{amsfonts} 
\usepackage{amssymb} 
\usepackage{xspace} 
\usepackage{latexsym} 
\usepackage{url} 
\usepackage{xspace} 
\usepackage{fancyvrb} 

\usepackage{graphics,color}              
\RequirePackage[latin1]{inputenc}   
 
\newenvironment{restate-proposition}[2][{}]{\noindent\textbf{Proposition~{#2}}\;\textbf{#1}\  
}{\vskip 1em} 
 
\newenvironment{restate-theorem}[2][{}]{\noindent\textbf{Theorem~{#2}}\;\textbf{#1}\  
}{\vskip 1em} 
 
\newenvironment{restate-corollary}[2][{}]{\noindent\textbf{Corollary~{#2}}\;\textbf{#1}\  
}{\vskip 1em}

\newcommand{\Proofitemb}[1]{\medskip \noindent {\bf #1\;}} 
\newcommand{\Proofitemfb}[1]{\noindent {\bf #1\;}} 
\newcommand{\Proofitem}[1]{\medskip \noindent $#1\;$} 
\newcommand{\Proofitemf}[1]{\noindent $#1\;$} 
\newcommand{\Defitem}[1]{\smallskip \noindent $#1\;$} 
 
\newcommand{\Defitemf}[1]{\noindent $#1\;$}

\makeatother 
 
\makeatletter 
\def\@ysproof[#1]{\@sproof{ #1}} 
\def\@sproof#1{\begin{trivlist}\item[]{\textit{Sketch of the proof#1.}}}

\makeatletter 
\def\@yproof[#1]{\@proof{ #1}} 
\def\@proof#1{\begin{trivlist}\item[]{\textit{Proof#1.}}}

\newcommand{\hbra}{\noindent\hbox to \textwidth{\leaders\hrule height1.8mm depth-1.5mm\hfill}} 
\newcommand{\hket}{\noindent\hbox to \textwidth{\leaders\hrule height0.3mm\hfill}} 
\newcommand{\ratio}{.3}

\newtheorem{theorem}{Theorem} 
\newtheorem{fact}[theorem]{Fact} 
\newtheorem{definition}[theorem]{Definition} 
\newtheorem{lemma}[theorem]{Lemma} 
 
\newtheorem{proposition}[theorem]{Proposition}

\newtheorem{remark}[theorem]{Remark}

\newcommand{\Proof}{\noindent {\sc Proof}. } 
 
\newcommand{\qed}{\hfill${\Box}$}

\newcommand{\Figbar}{{\center \rule{\hsize}{0.3mm}}}    
 
\newcommand{\cl}[1]{{\cal #1}}          

\newcommand{\ul}[1]{\underline{#1}}     
\newcommand{\ol}[1]{\overline{#1}}      
 
\newcommand{\arrow}{\rightarrow}        
\newcommand{\Alt}{ \mid\!\!\mid  } 
\newcommand{\isum}{\oplus} 
 
\newcommand{\infer}[2]{\begin{array}{c} #1 \\ \hline #2 \end{array}} 
 
\newcommand{\Or}{\vee}                  
\newcommand{\AND}{\wedge}               
 
\newcommand{\wbis}{\approx}             

\newcommand{\union}{\cup}               
\newcommand{\inter}{\cap}               
\newcommand{\Union}{\bigcup}            
\newcommand{\minus}{\backslash}         
\newcommand{\comp}{\circ}               
\newcommand{\set}[1]{\{#1\}}            
 
\newcommand{\dcl}{\downarrow}           
\newcommand{\ucl}{\uparrow}             
\newcommand{\nor}{\succeq}
\newcommand{\rl}[1]{\;{\cal #1}\;}             
\newcommand{\rel}[1]{{\cal #1}}         
 
\newcommand{\mand}{\mbox{ and }} 
\newcommand{\w}[1]{{\it #1}}    

\newcommand{\xst}[2]{\exists\, #1\;\: #2}

\newcommand{\s}[1]{{\sf #1}}    
\newcommand{\vc}[1]{{\bf #1}}

\newcommand{\act}[1]{\xrightarrow{#1}} 

\newcommand{\acteq}[1]{\stackrel{#1}{\leadsto}} 

\newcommand{\actI}[1]{\xrightarrow{#1}_{1}}

\newcommand{\actII}[1]{\xrightarrow{#1}_{2}}

 \newcommand{\wact}[1]{\stackrel{#1}{\Rightarrow}} 
\newcommand{\wactI}[1]{\stackrel{#1}{\Rightarrow_{1}}} 
\newcommand{\wactII}[1]{\stackrel{#1}{\Rightarrow_{2}}} 

\newcommand{\eval}{\Downarrow}

\newcommand{\susp}{\downarrow} 
\newcommand{\lsusp}{\Downarrow_L} 
\newcommand{\wsusp}{\Downarrow} 
\newcommand{\commits}{\searrow}

\newcommand{\spi}{S\pi} 
 
\newcommand{\bbis}{\approx_{B}} 
\newcommand{\cbis}{\approx_{C}} 
\newcommand{\lbis}{\approx}

\newcommand{\sbis}{\equiv} 
\newcommand{\emit}[2]{\ol{#1}#2}  
\newcommand{\present}[4]{#1(#2).#3,#4} 
\newcommand{\match}[4]{[#1=#2]#3,#4}       

\newcommand{\matchv}[4]{[#1 \unrhd #2]#3,#4}

\newcommand{\new}[2]{\nu #1 \ #2} 
\newcommand{\outact}[3]{\new{{\bf #1}}{\emit{#2}{#3}}} 
\newcommand{\real}{\makebox[5mm]{\,$\|\!-$}}
 
\begin{document} 
 
\title{Determinacy in a synchronous $\pi$-calculus 
\thanks{Work partially supported by ANR-06-SETI-010-02.} 
} 
\author{Roberto M. Amadio
 \qquad Mehdi Dogguy \\
        Universit\'e Paris Diderot, PPS, UMR-7126} 

\maketitle 

\begin{abstract}
The $\spi$-calculus is a {\em synchronous} $\pi$-calculus which is
based on the {\sc SL} model.  The latter is a relaxation of the {\sc
Esterel} model where the reaction to the {\em absence} of a signal
within an instant can only happen at the next instant.  In the present
work, we present and characterise a compositional semantics of the
$\spi$-calculus based on suitable notions of labelled transition
system and bisimulation. Based on this semantic framework, we explore
the notion of determinacy and the related one of (local) confluence.
\end{abstract}

\section{Introduction}
Let $P$ be a program that can repeatedly interact with its
environment.  A {\em derivative} of $P$ is a program to which $P$
reduces after a finite number of interactions with the environment.  A
program {\em terminates} if all its internal computations
terminate and it is {\em reactive} if all its derivatives are
guaranteed to terminate.  A program is {\em determinate} if after
any finite number of interactions with the environment the resulting
derivative is unique up to {\em semantic equivalence}.

Most conditions found in the literature that entail determinacy are
rather intuitive, however the formal statement of these conditions and the
proof that they indeed guarantee determinacy can be
rather intricate in particular in the presence of name mobility, as
available in a paradigmatic form in the $\pi$-calculus.

Our purpose here is to provide a streamlined theory of determinacy for
the {\em synchronous} $\pi$-calculus introduced in \cite{Amadio06}.
It seems appropriate to address these issues in a volume
dedicated to the memory of Gilles Kahn. First, Kahn networks \cite{Kahn74}
are a classic example of concurrent {\em and} deterministic systems.
Second, Kahn networks have largely inspired the research on {\em
synchronous} languages such as \textsc{Lustre} \cite{CPHP87} and, to a lesser
extent, \textsc{Esterel} \cite{BG92}.  
An intended side-effect of this work is to
illustrate how ideas introduced in concurrency theory well after Kahn
networks can be exploited to enlighten the study of determinacy in
concurrent systems.

Our technical approach will follow a process calculus tradition, namely:

\begin{enumerate}

\item We describe the interactions of a program with its environment
through a {\em labelled transition system} to which we associate
a compositional notion of {\em labelled bisimulation}.

\item We rely on this semantic framework, to introduce a notion of {\em determinacy} and a related notion of {\em confluence}.

\item We provide {\em local} confluence conditions that are easier to check
and that combined with {\em reactivity} turn out 
to be equivalent to determinacy.

\end{enumerate}

We briefly trace the path that has lead to this approach.
A systematic study of determinacy and confluence for CCS is available
in \cite{Milner89} where, roughly, the usual theory of rewriting is
generalised in two directions: first rewriting is labelled and second
diagrams commute up to semantic equivalence.  In
this context, a suitable formulation of Newman's lemma \cite{N42}, has
been given in \cite{GS96}.  The theory has been gradually extended
from CCS, to CCS with values, and finally to the $\pi$-calculus
\cite{PW97}. 

Calculi such as CCS and the $\pi$-calculus are designed to represent
{\em asynchronous} systems. On the other hand, the $\spi$-calculus is
designed to represent {\em synchronous} systems.  In these systems,
there is a notion of {\em instant} (or phase, or pulse, or round) and
at each instant each thread performs some actions and synchronizes
with all other threads. One may say that all threads proceed at the
same speed and it is in this specific sense that we will refer to {\em
synchrony} in this work.

In order to guarantee determinacy in the context of CCS {\em rendez-vous}
communication, it seems quite natural to restrict the calculus so that
interaction is {\em point-to-point}, {\em i.e.}, it involves exactly
one sender and one receiver.\footnote{Incidentally, this is also the
approach taken in Kahn networks but with an interaction mechanism
based on unbounded, ordered buffers.  It is not difficult to represent
unbounded, ordered buffers in a CCS with value passing and show that,
modulo this encoding, the determinacy of Kahn networks can be obtained
as a corollary of the theory of confluence developed in
\cite{Milner89}.}
In a synchronous framework, the introduction of {\em signal} based
communication offers an opportunity to move from point-to-point to a
more general multi-way interaction mechanism with multiple senders and/or
receivers, while preserving determinacy. In particular, this is the
approach taken in the {\sc Esterel} and SL \cite{BD95}
models.  The SL model can be regarded as a relaxation of the {\sc
Esterel} model where the reaction to the {\em absence} of
a signal within an instant can only happen at the next instant.  This
design choice avoids some paradoxical situations and simplifies the
implementation of the model. The SL model has gradually evolved into a
general purpose programming language for concurrent applications and
has been embedded in various programming environments such as
\textsc{C}, \textsc{Java}, \textsc{Scheme}, and \textsc{Caml} (see
\cite{boussinot:rc91,SchemeFT,MandelPouzetPPDP05}).  For
instance, the Reactive ML language \cite{MandelPouzetPPDP05} includes
a large fragment of the \textsc{Caml} language plus primitives to
generate signals and synchronise on them.  We should also mention that
related ideas have been developed by Saraswat et al. \cite{SJG96} in
the area of constraint programming.

The $\spi$-calculus  can be regarded as an
extension of the SL model where signals can carry values.  In this
extended framework, it is more problematic to have both concurrency
{\em and} determinacy.  Nowadays, this question 
is frequently considered when designing various kind of synchronous 
programming languages
(see, {\em e.g.}, \cite{MandelPouzetPPDP05,ET06}). 
As we already mentioned, our
purpose here is to address the question with the tool-box of process
calculi following the work for CCS and the $\pi$-calculus quoted
above.  In this respect, it is worth stressing 
a few interesting variations that arise when
moving from the `asynchronous' $\pi$-calculus to 
the `synchronous' $\spi$-calculus.  First, we
have already pointed-out that there is an opportunity to move from a
point-to-point to a multi-way interaction mechanism while preserving
determinacy.  Second, the notion of confluence and determinacy happen
to coincide while in the asynchronous context confluence is a
strengthening of determinacy which has better compositionality
properties.  Third, reactivity appears to be a reasonable property to
require of a synchronous system, the goal being just to avoid
instantaneous loops, {\em i.e.}, loops that take 
no time.\footnote{The situation is different in asynchronous 
systems where reactivity
is a more demanding property.
For instance, \cite{GS96} notes: ``{\em As soon
as a protocol internally consists in some kind of correction mechanism
(e.g., retransmission in a data link protocol) the specification of
that protocol will contain a $\tau$-loop}''. }

The rest of the paper is structured as follows.
In section \ref{section-spi}, we introduce the $\spi$-calculus,
in section \ref{semantics-sec}, we define its semantics based on
a standard notion of labelled bisimulation on a 
(non-standard) labelled transition system
and we show that the bisimulation is preserved by static contexts, 
in section \ref{charact-sec} we provide alternative characterisations
of the notion of labelled bisimulation we have introduced,
in section \ref{det-conf-sec}, we develop the concepts of determinacy
and (local) confluence.
Familiarity with the $\pi$-calculus \cite{MPW92,SW01}, the notions of determinacy
and confluence presented in \cite{Milner89}, and synchronous languages
of the {\sc Esterel} family \cite{BG92,BD95} is assumed.

\section{Introduction to the $\spi$-calculus}\label{section-spi}
We introduce the syntax of the $\spi$-calculus along with 
an informal comparison with the $\pi$-calculus and a  programming example.

\subsection{Programs}\label{programs}
Programs $P,Q,\ldots$ in the $\spi$-calculus
are defined as follows:
\[
\begin{array}{ll}
P &::= 0 \Alt A(\vc{e}) \Alt \emit{s}{e} \Alt
\present{s}{x}{P}{K} 
\Alt \match{s_1}{s_2}{P_1}{P_2}
\Alt \matchv{u}{p}{P_1}{P_2}
\Alt \new{s}{P}
\Alt P_1\mid P_2 \\
K &::=A(\vc{r})
\end{array}
\]
We use the notation $\vc{m}$ for a vector $m_1,\ldots,m_n$, $n\geq 0$.
The informal behaviour of programs follows.
$0$ is the terminated thread. $A(\vc{e})$ is a (tail) recursive call
of a thread identifier $A$ with a vector $\vc{e}$ of expressions as argument;
as usual the thread identifier $A$ is defined by a unique equation
$A(\vc{x})=P$ such that the free variables of $P$ occur in $\vc{x}$.
$\emit{s}{e}$ evaluates the expression $e$ and emits its value on the
signal $s$.  $\present{s}{x}{P}{K}$ is the {\em present} statement
which is the fundamental operator of the SL model. If the values
$v_1,\ldots,v_n$ have been emitted on the signal $s$ 
then $\present{s}{x}{P}{K}$ evolves
non-deterministically into $[v_i/x]P$ for some $v_i$ ($[\_/\_]$ is our
notation for substitution).  On the other
hand, if no value is emitted then the continuation $K$ is evaluated at
the end of the instant.  $\match{s_1}{s_2}{P_1}{P_2}$ is the usual
matching function of the $\pi$-calculus 
that runs $P_1$ if $s_1$ equals $s_2$ and $P_2$, otherwise.
Here both $s_1$ and $s_2$ are free.
$\matchv{u}{p}{P_1}{P_2}$, matches $u$ against the pattern $p$.
We assume $u$ is either a variable $x$ or a value $v$ and $p$ has
the shape $\s{c}(\vc{x})$, where $\s{c}$ is a constructor and 
$\vc{x}$ is a vector of distinct variables.  We also assume that if $u$ is a
variable $x$ then $x$ does not occur free in $P_{1}$.
At run time, $u$ is always a {\em value} and we 
run $\theta P_1$ if $\theta=\w{match}(u,p)$ is the 
substitution matching $u$ against $p$, and $P_2$ if such 
substitution does not exist (written $\w{match}(u,p)\ucl$).
Note that as usual the variables occurring in the pattern $p$
(including signal names) are bound in $P_1$.
$\new{s}{P}$ creates a new signal name $s$ and runs $P$.
$(P_1\mid P_2)$ runs in parallel $P_1$ and $P_2$.  A continuation $K$
is simply a recursive call whose arguments are either expressions
or values associated with signals at the end of the instant in
a sense that we explain below. We will also write 
$\s{pause}.K$ for $\new{s}{\present{s}{x}{0}{K}}$ with $s$ not free in $K$. 
This is the program that waits till the end of the instant and then
evaluates $K$.

\subsection{Expressions}\label{expressions}
The definition of programs relies on the following syntactic categories:
\[
\begin{array}{lll}
\w{Sig} &::= s \Alt t \Alt \cdots  &\mbox{(signal names)} \\
\w{Var} &::= \w{Sig} \Alt x \Alt y \Alt z \Alt \cdots   &\mbox{(variables)} \\
\w{Cnst} &::= \s{*} \Alt \s{nil} \Alt \s{cons} \Alt \s{c} \Alt \s{d} \Alt\cdots &\mbox{(constructors)} \\
\w{Val} &::= \w{Sig} \Alt \w{Cnst}(\w{Val},\ldots,\w{Val})
&\mbox{(values $v,v',\ldots$)}\\
\w{Pat} &::=  \w{Cnst}(\w{Var},\ldots,\w{Var})
&\mbox{(patterns $p,p',\ldots$)} \\
\w{Fun} &::=f \Alt g \Alt \cdots &\mbox{(first-order function
  symbols)} \\
\w{Exp} &::= \w{Var} \Alt \w{Cnst}(\w{Exp},\ldots,\w{Exp}) \Alt 
                          \w{Fun}(\w{Exp},\ldots,\w{Exp}) 
&\mbox{(expressions $e,e',\ldots$)} \\
\w{Rexp} &::= {!\w{Sig}} \Alt \w{Var} \Alt 
\w{Cnst}(\w{Rexp},\ldots,\w{Rexp}) \Alt \\ &\quad \w{Fun}(\w{Rexp},\ldots,\w{Rexp})
&\mbox{(exp. with deref. $r,r',\ldots$)} 
\end{array}
\]
As in the $\pi$-calculus, signal names stand both for
signal constants as generated by the $\nu$ operator and signal
variables as in the formal parameter of the present operator.
Variables $\w{Var}$ include signal names as well as variables of other
types.  Constructors $\w{Cnst}$ include $\s{*}$, $\s{nil}$, and $\s{cons}$.
Values $\w{Val}$ are terms built out of constructors and signal names.
Patterns $\w{Pat}$ are terms built out of constructors and variables
(including signal names).  
If $P, p$ are a program and a pattern then  we denote
with $\w{fn}(P), \w{fn}(p)$ the set of free signal names
occurring in them, respectively. We also use $\w{FV}(P), \w{FV}(p)$
to denote the set of free variables (including signal names).
We assume first-order function symbols $f,g,\ldots$  and
an evaluation relation $\eval$ such that for every
function symbol $f$ and values $v_1,\ldots,v_n$ of suitable type
there is a unique value $v$ such that $f(v_1,\ldots,v_n)\eval v$ and
$\w{fn}(v)\subseteq \Union_{i=1,\ldots,n}\w{fn}(v_{i})$.
Expressions $\w{Exp}$ are terms built out of variables, constructors,
and function symbols. The evaluation relation $\eval$ is extended
in a standard way to expressions whose only free variables are
signal names.
Finally, $\w{Rexp}$ are expressions that may include 
the value associated with a signal $s$ at the
end of the instant (which is written $!s$, following the ML notation
for dereferenciation). Intuitively, this value
is a {\em list of values} representing the {\em set of values} emitted on 
the signal during the instant.

\subsection{Typing}
Types include  the basic type $1$ inhabited by the constant $*$ and, 
assuming $\sigma$ is a type,  the type $\w{Sig}(\sigma)$ of signals carrying
values of type $\sigma$, and the type $\w{List}(\sigma)$ of lists of values of
type $\sigma$ with constructors \s{nil} and \s{cons}.
In the examples, it will be convenient to abbreviate 
$\s{cons}(v_{1},\ldots,\s{cons}(v_{n},\s{nil})\ldots)$ with
$[v_{1};\ldots;v_{n}]$.
$1$ and $\w{List}(\sigma)$ are examples of {\em inductive types}. More
inductive types (booleans, numbers, trees,$\ldots$) 
can be added along with more constructors.  
We assume that variables (including signals), constructor
symbols, and thread identifiers come with their (first-order) types. 
For instance, a function symbols $f$ may have a type
$(\sigma_1,\sigma_2)\arrow \sigma$ meaning that it waits two arguments of
type $\sigma_1$ and $\sigma_2$ respectively and returns a value of type 
$\sigma$.
It is straightforward to define when a program is well-typed.  We just
point-out that if a signal name $s$ has type $\w{Sig}(\sigma)$ then
its dereferenced value $!s$ has type $\w{List}(\sigma)$. In the
following, we will tacitly assume that we are handling well typed
programs, expressions, substitutions,$\ldots$

\subsection{Comparison with the $\pi$-calculus}
The syntax of the $\spi$-calculus is similar to the one of
the $\pi$-calculus, however there are some important {\em semantic}
differences that we highlight in the following simple example.
Assume $v_1\neq v_2$ are two distinct values and consider the
following program in $\spi$:
\[
\begin{array}{l}
P=\nu \ s_1,s_2 \ 
(\quad \emit{s_{1}}{v_{1}} \quad \mid \quad 
 \emit{s_{1}}{v_{2}} \quad \mid  \quad 
 s_1(x). \ (s_1(y). \  (s_2(z). \ A(x,y) \  \ul{,B(!s_1)}) \quad
 \ul{,0}) \quad \ul{,0} \quad )
\end{array}
\]
If we forget about the underlined parts and we regard $s_1,s_2$ as
{\em channel names} then $P$ could also be viewed as a $\pi$-calculus
process. In this case, $P$ would reduce to 
\[
P_1 =  \new{s_1,s_2}{(s_2(z).A(\theta(x),\theta(y))}
\]
where $\theta$ is a substitution such that
$\theta(x),\theta(y)\in \set{v_1,v_2}$ and $\theta(x)\neq \theta(y)$.
In $\spi$, {\em signals persist within the instant} and 
$P$ reduces to 
\[
P_2 = \new{s_1,s_2}{(\emit{s_{1}}{v_{1}} \mid 
 \emit{s_{1}}{v_{2}} \mid (s_2(z).A(\theta(x),\theta(y))\ul{,B(!s_1)}))}
\]
where  $\theta(x),\theta(y)\in \set{v_1,v_2}$.
What happens next? In the $\pi$-calculus, $P_1$ is 
{\em deadlocked} and no further computation is possible.
In the $\spi$-calculus, 
the fact that no further computation is possible in $P_2$ is
detected and marks the {\em end of the current instant}. Then
an additional computation represented by the relation $\act{N}$ 
moves $P_2$ to the following instant:
\[
P_2 \act{N} P'_2 =  \new{s_1,s_2}{B(v)}
\]
where $v \in  \set{[v_1;v_2],[v_2;v_1]}$.
Thus at the end of the instant, a dereferenced signal such as $!s_{1}$
becomes a list of (distinct) values emitted on $s_1$ during the
instant and then all signals are reset.

\subsection{A programming example}\label{programming-example}\label{server-ex}
We introduce a programming example to illustrate the
kind of synchronous programming that can be represented in the
$\spi$-calculus. 
We describe first a `server' handling a list of requests
emitted in the previous instant on the signal $s$. 
For each request of the shape $\s{req}(s',x)$, 
it provides an answer which is a function of $x$ along the signal $s'$.
\[
    \begin{array}{lcl}
      \w{Server}(s)&=&{\tt{pause}}.\w{Handle}(s,!s)\\
      \w{Handle}(s,\ell)&=&
      \matchv{\ell}{\s{req}(s',x)::\ell'}
      {(\emit{s'}{f(x)} \mid \w{Handle}(s,\ell'))}
      {\w{Server}(s)}~.
    \end{array}
\]
The programming of a client that issues a request $x$ on signal $s$
and returns the reply on signal $t$ could be the following:
\[
\begin{array}{lcl}
\w{Client}(x,s,t) &=&\nu s' \ (\emit{s}{\s{req}(s',x)} \mid \s{pause}.s'(x).\emit{t}{x},0)~.
\end{array}
\]

\section{Semantics of the $\spi$-calculus}\label{semantics-sec}
In this section, we define the semantics of the $\spi$-calculus by
a `standard'  notion of labelled bisimulation on a `non-standard' 
labelled transition system and we show that  labelled bisimulation 
is preserved by `static' contexts.
A distinct notion of labelled bisimulation for the $\spi$-calculus has
already been studied in \cite{Amadio06} and the following section
\ref{charact-sec} will show that the two notions are (almost) the
same.
A significant advantage of the presentation of labelled bisimulation we discuss
here is that in the `bisimulation game'
all actions are treated in the same way.  This allows
allows for a considerable simplification of the diagram chasing
arguments that are needed in the study of determinacy and confluence
in section \ref{det-conf-sec}.

\subsection{Actions}\label{action-sec}
The actions of the forthcoming labelled transition system 
are classified in the following categories:
\[
\begin{array}{lll}
\w{act} 
&::= \alpha \Alt \w{aux}              
&(\mbox{actions}) \\

\alpha  
&::= \tau \Alt \outact{t}{s}{v} \Alt sv \Alt N 
&(\mbox{relevant actions})\\

\w{aux} 
&::= s?v \Alt (E,V) 
&(\mbox{auxiliary actions})\\

\mu     
&::= \tau \Alt \outact{t}{s}{v} \Alt s?v 
&(\mbox{nested actions})
\end{array}
\]
The category $\w{act}$ is partitioned into relevant actions
and auxiliary actions. 

The {\em relevant actions} are those that are actually considered in the
bisimulation game. They consist of: (i) an internal action $\tau$, (ii)
an emission action $\outact{t}{s}{v}$ where it is assumed that the
signal names $\vc{t}$ are distinct, occur in $v$, and differ from $s$,
(iii) an input action $sv$, and (iv) an action $N$ (for {\em Next})
that marks the move from the current to the next instant.

The {\em auxiliary actions} consist of an input action $s?v$ which is
coupled with an emission action in order to compute a $\tau$ action and
an action $(E,V)$ which is just needed to compute an action $N$.
The latter is an action that can occur exactly when
the program cannot perform $\tau$ actions and it amounts (i) to collect
in lists the set of values emitted on every signal, (ii) to reset all
signals, and (iii) to initialise the continuation $K$
for each present statement of the shape $s(x).P,K$.

In order to formalise these three steps we need to introduce 
some notation.
Let $E$ vary over functions from signal names
to finite sets of values. Denote with $\emptyset$
the function that associates the empty set with every
signal name,  with $[M/s]$ the function that associates 
the set $M$ with the signal name $s$ and the empty set
with all the other signal names, and with $\union$ the 
union of functions defined point-wise.

We represent a set of values as a list of the
values contained in the set. More precisely,
we write $v \real M$ and say that $v$ {\em represents} $M$ 
if $M=\set{v_1,\ldots,v_n}$ and
$v=[v_{\pi(1)};\ldots; v_{\pi(n)}]$ for
some permutation $\pi$ over $\set{1,\ldots,n}$.
Suppose $V$ is a function from signal names to lists of values.
We write $V\real E$ if $V(s)\real E(s)$ for every signal name $s$.
We also write $\w{dom}(V)$ for 
$\set{s \mid V(s)\neq []}$.
If $K$ is a continuation, {\em i.e.}, a recursive call $A(\vc{r})$,
then $V(K)$ is obtained from $K$ by replacing
each occurrence $!s$ of a dereferenced signal with the associated
value $V(s)$. We denote with $V[\ell/s]$ the function that behaves as
$V$ except on $s$ where $V[\ell/s](s)=\ell$.

With these conventions, a transition
$P\act{(E,V)} P'$ intuitively means that 
(1) $P$ is suspended,
(2) $P$ emits exactly the values specified by $E$, and
(3) the behaviour of $P$ in the following instant is $P'$ and depends
on  $V$. 
It is convenient to compute these transitions on programs where
all name generations are lifted at top level. We write $P \nor Q$
if we can obtain $Q$ from $P$ by repeatedly  transforming, for instance,
a subprogram $\nu s P' \mid P''$ into $\nu s (P'\mid P'')$ 
where $s\notin \w{fn}(P'')$.

Finally, the {\em nested actions} $\mu,\mu',\ldots$ are certain
actions (either relevant or auxiliary) that can be  produced by a sub-program 
and that we need to propagate to the top level.

\subsection{Labelled transition system}
The labelled transition system 
is defined in table \ref{Lts} where rules apply to
programs whose only free variables are signal names and with
standard conventions on the renaming of bound names.
As usual, one can rename bound variables, and 
the symmetric rules for $(\w{par})$ and $(\w{synch})$
are omitted.
The first $12$ rules from $(\w{out})$ to $(\nu_{\w{ex}})$
are quite close to those of a polyadic $\pi$-calculus with
asynchronous communication (see \cite{HT92,HY95,ACS98})
with the following exception: rule $(\w{out})$ models the fact
that the emission of a value on a signal {\em persists} within
the instant. The last $5$ rules from $(0)$ to $(\w{next})$
are quite specific of the $\spi$-calculus
and determine how the computation is carried on at the end
of the instant (cf. discussion in \ref{action-sec}).

The relevant actions different from $\tau$, model the
possible interactions of a program with its environment.
Then the notion of reactivity can be formalised as follows.

\begin{definition}[derivative] 
A derivative of a program $P$ is a program $Q$ such that  
\[ 
P \act{\alpha_{1}} \cdots \act{\alpha_{n}} Q, \qquad\mbox{where: }n\geq 0 ~.
\] 
\end{definition}

\begin{definition}[reactivity] \label{react-def}
We say that a program $P$ is reactive, if for every derivative $Q$ 
every $\tau$-reduction sequence terminates.  
\end{definition}

\begin{table}
{\small
\[
\begin{array}{cc}

(\w{out})\quad\infer{e\eval v}{\emit{s}{e} \act{\emit{s}{v}} \emit{s}{e}}

&(\w{in}_{\w{aux}})\quad
\infer{~}{\present{s}{x}{P}{K} \act{s?v} [v/x]P}
\\ \\

(\w{in})\quad \infer{~}{P\act{sv} (P \mid \emit{s}{v})} 

&(\w{rec})\quad \infer{A(\vc{x})=P,\quad \vc{e}\eval \vc{v}}{A(\vc{e})\act{\tau}[\vc{v}/\vc{x}]P}  \\ \\

(=_{1}^{\w{sig}})\quad \infer{~}{\match{s}{s}{P_1}{P_2}\act{\tau} P_1}

&(=_{2}^{\w{sig}})\quad 
\infer{s_1\neq s_2}{\match{s_1}{s_2}{P_1}{P_2}\act{\tau} P_2} \\ \\

(=_{1}^{\w{ind}})\quad
\infer{\w{match}(v,p)=\theta}{\matchv{v}{p}{P_1}{P_2}\act{\tau} \theta P_1}

&(=_{1}^{\w{ind}})\quad
\infer{\w{match}(v,p)=\ucl}{\matchv{v}{p}{P_1}{P_2} \act{\tau} P_2}  \\ \\

(\w{comp})\quad
\infer{P_1\act{\mu} P'_1 \quad \w{bn}(\mu)\inter\w{fn}(P_2)=\emptyset}
{P_1\mid P_2\act{\mu} P'_1\mid P_2}

&(\w{synch})\quad
\infer{\begin{array}{c}
P_1 \act{\outact{t}{s}{v}} P'_1\quad P_2 \act{s?v}P'_2\\
\set{\vc{t}}\inter \w{fn}(P_2)=\emptyset\end{array}}
{P_1\mid P_2 \act{\tau} \new{\vc{t}}{(P'_1\mid P'_2)}} \\ \\

(\nu)\quad \infer{P\act{\mu} P' \quad t\notin n(\mu)}
{\new{t}{P}\act{\mu} \new{t}{P'}}

&(\nu_{\w{ex}})\quad
\infer{P\act{\outact{t}{s}{v}} P'\quad t'\neq s\quad t'\in n(v)\minus \set{\vc{t}}}
{\new{t'}{P}\act{(\nu t',\vc{t})\emit{s}{v}} P'}  \\ \\

(0)\quad
\infer{~}
{0 \act{\emptyset,V} 0}

&(\w{reset})\quad
\infer{e\eval v \quad v \mbox{ occurs in }V(s)}
{\emit{s}{e} \act{[\set{v}/s],V} 0} \\ \\

(\w{cont})\quad
\infer{s\notin \w{dom}(V)}
{s(x).P,K \act{\emptyset,V} V(K)} 

&(\w{par})\quad
\infer{P_i \act{E_{i},V} P'_i \quad i=1,2 }
{(P_1\mid P_2) \act{E_1\union E_2,V} (P'_1\mid P'_2)} \\ \\

(\w{next})\quad
\infer{P\nor \nu \vc{s}\  P' \quad P' \act{E,V} P''\quad V\real E}
{P\act{N} \nu \vc{s} \ P''}

\end{array}
\]}
\caption{Labelled transition system}\label{Lts}
\end{table}

\subsection{A compositional labelled bisimulation}
We introduce first a rather standard notion of (weak)
labelled bisimulation. We define $\wact{\alpha}$ as:
{\small
\[
\wact{\alpha} = 
\left\{
\begin{array}{ll}
(\act{\tau})^* 
&\mbox{if }\alpha=\tau \\

(\wact{\tau})\comp(\act{N})
&\mbox{if }\alpha=N \\

(\wact{\tau})\comp (\act{\alpha})\comp (\wact{\tau})
&\mbox{otherwise}
\end{array} \right.
\]}
This is the standard definition except that we insist on {\em not}
having internal reductions after an $N$ action.
Intuitively, we assume that an observer can control the execution
of programs so as to be able to test them
at the very beginning of each instant.\footnote{
This decision entails that,
{\em e.g.}, we distinguish the programs $P$ and $Q$ defined as follows:
$P = \s{pause}.(\ol{s}_{1} \isum \ol{s}_{2})$,
$Q = \nu s \ (\s{pause}.A(!s) \mid \emit{s}{0} \mid \emit{s}{1})$,
 where
$A(x) = \matchv{x}{[0;1]}{(\ol{s}_{1}\isum \ol{s}_{2})}{\ol{s}_{1}}$,
and  $\isum$, $0$, and $1$ are abbreviations for 
an internal choice and for two distinct constants, respectively
(these concepts can be easily coded in the $\spi$-calculus).
On the other hand, $P$ and $Q$ would be equivalent if we defined
$\wact{N}$ as $\wact{\tau} \comp \act{N} \comp \wact{\tau}$.}
We write $P\act{\alpha} \cdot$ for $\xst{P'}{(P\act{\alpha}P')}$.

\begin{definition}[labelled bisimulation]\label{def-bis}
A symmetric relation $\rel{R}$ on programs is a labelled bisimulation
if 
\[
\infer{P\rl{R} Q,\quad P\act{\alpha} P', \quad \w{bn}(\alpha)\inter \w{fn}(Q)=\emptyset}
{\xst{Q'}{( \ Q\wact{\alpha} Q',\qquad P'\rl{R} Q' \ )}}
\]
\smallskip\noindent
We denote with $\wbis$ the largest labelled bisimulation.
\end{definition}

The standard variation where one considers weak reduction in the
hypothesis ($P\wact{\alpha} P'$ rather than $P\act{\alpha} P'$)
leads to the same relation. Also, relying on this
variation, one can show that the concept of bisimulation up to
bisimulation makes sense, {\em i.e.}, a bisimulation up
to bisimulation is indeed contained in the largest bisimulation.
An important property of labelled bisimulation is that
it is preserved by static contexts. The proof of this fact
follows \cite{Amadio06} and it is presented in appendix \ref{compose-bis-proof}.

\begin{definition} \label{static-cxt-def}
A static context $C$ is defined as follows:
\begin{equation}\label{contexts}
C::= [~] \Alt C\mid P \Alt \new{s}{C}
\end{equation}
\end{definition}

\begin{theorem}[compositionality of labelled bisimulation]\label{compose-bis-thm}
If $P\wbis Q$ and $C$ is a static context then $C[P]\wbis C[Q]$.
\end{theorem}

\section{Characterisations of labelled bisimulation}\label{charact-sec}
The labelled transition system presented in table \ref{Lts}
embodies a number of technical choices which might not appear
so natural at first sight. To justify these choices, 
it is therefore interesting to look for 
alternative characterisations of the induced bisimulation equivalence.
To this end we recall the notion of {\em contextual} bisimulation introduced
in \cite{Amadio06}.

\begin{definition}\label{wl-susp-def}
We write: 
\[
\begin{array}{llll}

P\susp   &\mbox{if}&\neg (\ P \act{\tau} \cdot \ )
&\mbox{(suspension)} \\ 

P \wsusp &\mbox{if} &\xst{P'}{( \ P\wact{\tau} P' \mand P'\susp \ )}
&\mbox{(weak suspension)} \\ 
P\lsusp 
&\mbox{if} &\xst{P'}{( \ P\mid P' \ ) \wsusp}
&\mbox{(L-suspension)}
\end{array}
\]
\end{definition}

Obviously, $P\susp$ implies $P\wsusp$ which in turn implies $P\lsusp$
and none of these implications can be reversed (see
\cite{Amadio06}).  Also note that all the derivatives of a reactive
program enjoy the weak suspension property.

\begin{definition}[commitment]\label{commit-def}
We write $P\commits \ol{s}$ if
$P\act{\outact{t}{s}{v}}\cdot$ and say that
$P$ commits to emit on $s$.
\end{definition}

\begin{definition}[barbed bisimulation]\label{barbed-bisimulation}
A symmetric relation $\rel{R}$ on programs is a barbed bisimulation if
whenever $P\rl{R} Q$ the following holds:

\Defitem{(B1)}
If $P\act{\tau} P'$ then
$\xst{Q'}{(Q\wact{\tau} Q' \mand P'\rl{R}Q')}$.

\Defitem{(B2)}
If $P\commits \ol{s}$ and $P\lsusp$ then
$\xst{Q'}{(Q\wact{\tau} Q', Q'\commits \ol{s}, 
\mand P \rl{R}Q')}$.

\Defitem{(B3)}
If $P\susp$  and $P\act{N} P''$ then 
$\xst{Q', Q''}{(Q\wact{\tau} Q', Q'\susp, P \rl{R}Q',
Q'\act{N} Q'',  \mand 
P'' \rl{R}  Q'')}$.

\smallskip\noindent
We denote with $\bbis$ the largest barbed bisimulation.
\end{definition}

\begin{definition}[contextual bisimulation]\label{cont-bis-def}
A symmetric relation $\rel{R}$ on programs is a contextual bisimulation if
it is a barbed bisimulation (conditions $(B1-3)$) and moreover 
whenever $P\rl{R} Q$ then

\Defitem{(C1)} $C[P] \rl{R} C[Q]$, for any static context $C$.

\smallskip\noindent
We denote with $\cbis$ the largest contextual barbed bisimulation.
\end{definition}

We arrive at the announced characterisation of the labelled
bisimulation.  

\begin{theorem}[characterisation of labelled bisimulation]\label{char-bis}
If $P,Q$ are reactive programs then $P\wbis Q$ if and only if $P\cbis Q$.
\end{theorem}

The proof of this result takes several steps summarised
in Table \ref{equiv-bis} which provides $3$ {\em equivalent} 
formulations of the labelled bisimulation $\wbis$. 
\begin{table}
{\footnotesize
\[
\begin{array}{|l|c|l|c|}

\hline
&\mbox{Labelled transition systems}

&&\mbox{Bisimulation game} \\\hline

&&&\\

(\actI{\alpha})

&\begin{array}{c}
\mbox{Rule }(\w{in}_{\w{aux}})\mbox{ replaced by} \\
\begin{array}{c}
(\w{in}^{1}_{\w{aux}}) \quad
\infer{~}
{s(x).P,K \act{s?v} [v/x]P\mid \emit{s}{v}}
\end{array}
\end{array}

&(\wbis_1)
&\mbox{As in definition }\ref{def-bis} \\ 

&&&\\\hline

&&&\\

(\actII{\alpha})

&\begin{array}{c}
\mbox{Rule }(\w{in})\mbox{ removed and} \\
\mbox{action }s?v\mbox{ replaced by }sv
\end{array}

&(\wbis_2)
&\begin{array}{c}
\mbox{As above if }\alpha\neq sv. \mbox{ Require:} \\
(\w{Inp})
\quad \infer{P \rl{R} Q}
{(P\mid \emit{s}{v})\rl{R} (Q\mid \emit{s}{v})}
\end{array} 
\\ 
&&&\\\hline

&&&\\

&\mbox{As above}

&(\wbis_3)

&\begin{array}{c}
\mbox{As above if }\alpha \neq sv. \mbox{ Replace }\w{(Inp)} 
\mbox{ with }: \\
\infer{P \rl{R} Q,\qquad  P\actII{sv} P'}
{\begin{array}{l}
 \exists Q'\ ( \ Q\wactII{sv} Q' \AND P' \rl{R} Q') \Or \\
 \qquad (Q\wactII{\tau} Q' \AND P' \rl{R}(Q'\mid \emit{s}{v}) \ )
\end{array}} \\\\

\mbox{and for }\alpha=N\mbox{ require:} \\

\infer{\begin{array}{c} P \rl{R} Q,\ (P\mid S)\act{N} P',
\\
S=\emit{s}_{1}v_{1}\mid \cdots \mid \emit{s}_{n}v_{n}\end{array}}
{\begin{array}{l}
 \exists Q',Q''\ ( \ (Q\mid S) \wactII{\tau} Q'',\quad (P\mid S) \rl{R} Q'',\\
 \qquad\qquad       Q''\actII{N} Q',\quad P'\rl{R}Q' \ )
\end{array}}
\end{array}
\\ 

&&&\\\hline
\end{array}
\]}
\caption{Equivalent formulations of labelled bisimulation}\label{equiv-bis}
\end{table}
In \cite{Amadio06}, the contextual bisimulation in definition 
\ref{cont-bis-def} is characterised as a variant of the
bisimulation $\wbis_3$ where the condition for the output
is formulated as follows:
\[
\infer{P\rl{R} Q,\qquad P\lsusp,\qquad P\actII{\outact{t}{s}{v}}
  P',\qquad \set{\vc{t}}\inter \w{fn}(Q)=\emptyset}
{Q\wactII{\outact{t}{s}{v}} Q',\quad P'\rl{R} Q'}
\]
Clearly, if $P$ is a reactive program then $P\lsusp$.
Also note that the definition \ref{react-def} of reactive program refers to
the labelled transition system \ref{Lts} for which it holds that 
$P\act{sv} (P\mid \emit{s}{v})$.
Therefore,  if $P$ is reactive then $(P\mid \emit{s}{v})$ is reactive
too and if we start comparing two reactive programs then all
programs that have to be considered in the bisimulation game will be
reactive too. 
This means that on reactive programs the condition $P\lsusp$ is 
always satisfied and therefore that the bisimulation $\wbis_3$ coincides
with the labelled bisimulation considered in
\cite{Amadio06}.\footnote{On 
non-reactive programs, labelled bisimulation makes more distinctions
than contextual bisimulation. For instance, the latter identifies
all the programs that do not L-suspend.}

\begin{remark}[on determinacy and divergence]
One may notice that the notions of labelled
bisimulation and contextual bisimulation we have adopted are only {\em
partially} sensitive to divergence.  Let $\Omega=\tau.\Omega$ be a
looping program.  Then $\Omega \not\wbis_{C} 0$ since $0$ may suspend
while $\Omega$ may not.
On the other hand, consider a program such as $A=\tau.A\isum \tau.0$.
Then $A\wbis 0$ and therefore $A\wbis_C 0$ and we are lead to conclude
that $A$ is a {\em determinate} program.  However, one may also argue
that $A$ is {\em not} determinate since it may either suspend or loop.
In other words, determinacy depends on the notion of semantic
equivalence we adopt. If the latter is not sensitive enough to
divergence then the resulting notion of determinacy should be regarded
as a {\em partial} property of programs, {\em i.e.}, it holds provided
programs terminate.
In practice, these distinctions do not seem very important because, as 
we have already argued, {\em reactivity} is a property one should
always require of synchronous programs and once reactivity is in place
the distinctions disappear.
\end{remark}

\section{Determinacy and (local) confluence}\label{det-conf-sec}
In this section, we develop the notions of determinacy
and confluence for the $\spi$-calculus which turn out to
coincide. Moreover, we note that for reactive
programs a simple property of local confluence suffices to ensure
determinacy. 

We denote with $\epsilon$ the empty sequence and with
$s=\alpha_1\cdots \alpha_n$ a finite sequence (possibly empty) of
actions different from $\tau$. We define:
\[
\wact{s} = \left\{
\begin{array}{ll}
\wact{\tau} &\mbox{if }s=\epsilon \\
\wact{\alpha_{1}} \cdots \wact{\alpha_{n}}
&\mbox{if }s=\alpha_1\cdots\alpha_n
\end{array}\right.
\]
Thus $s$ denotes a finite (possibly empty) sequence
of interactions with the environment. 
Following \cite{Milner89}, a program is considered
determinate if performing twice the same sequence of interactions leads
to the same program up to semantic equivalence.

\begin{definition}[determinacy] \label{det-def}
We say that a program $P$ is determinate if for every sequence $s$,
if $P\wact{s} P_i$ for $i=1,2$ then $P_1 \wbis P_2$. 
\end{definition} 

Determinacy implies $\tau$-inertness which is defined as follows.

\begin{definition}[$\tau$-inertness]
A program is $\tau$-inert if for all its derivatives $Q$,
$Q\act{\tau} Q'$ implies $Q\wbis Q'$. 
\end{definition} 

Next, we turn to the notion of confluence. To this end, we introduce
first the notions of action compatibility and action residual.

\begin{definition}[action compatibility]
The {\em compatibility predicate}  $\dcl$  is
defined as the least reflexive and symmetric binary relation
on actions such that $\alpha \dcl \beta$ implies that
either $\alpha,\beta\neq N$ or $\alpha=\beta=N$.
\end{definition}

In other words, the action $N$ is only compatible with itself while
any action different from $N$ is compatible with any other action
different from $N$.\footnote{The reader familiar with \cite{PW97}
will notice that, unlike in the $\pi$-calculus with
{\em rendez-vous} communication, we do not restrict the compatibility
relation on input actions. This is because of the particular form of 
the input action  in the labelled transition system in table \ref{Lts}
where the input action does not actually force a program to perform
an input. We expect that a similar situation would arise in the $\pi$-calculus
with asynchronous communication.}
Intuitively, confluence is about the possibility
of commuting actions that happen in the {\em same instant}. 
To make this precise we also need to introduce a notion 
of action residual $\alpha\minus \beta$ which specifies
what remains of the action $\alpha$ once the action $\beta$
is performed.

\begin{definition}[action residual]
 The residual operation $\alpha\minus \beta$ on actions is only defined if
 $\alpha\dcl \beta$ and in this case it satisfies:
\[
\alpha \minus \beta = \left\{
\begin{array}{ll}
\tau   &\mbox{if }\alpha=\beta \\

\nu \vc{t}\minus \vc{t'}\emit{s}{v} 
&\mbox{if }\alpha= \nu \vc{t}\ \emit{s}{v} \mbox{ and }\beta=\nu
\vc{t'}\emit{s'}{v'} \\

\alpha &\mbox{otherwise}
\end{array}
\right.
\]
\end{definition}

Confluence is then about closing diagrams of 
compatible actions up to residuals and semantic equivalence.

\begin{definition}[confluence] 
We say that a program $P$ is confluent, if for all its derivatives $Q$: 
\[ 
\infer{Q\wact{\alpha} Q_1,\quad Q\wact{\beta} Q_2,\quad \alpha \dcl \beta} 
{\xst{Q_3,Q_4}{( \ Q_1 \wact{\beta\minus\alpha} Q_3,\quad  
 Q_2 \wact{\alpha\minus\beta} Q_4, \quad 
 Q_3\wbis Q_4 \ )}} 
\] 
\end{definition} 

It often turns out that the following weaker notion of {\em local} confluence
is much easier to establish.

\begin{definition}[local confluence] 
We say that a program is locally confluent, if for all its derivatives $Q$: 
\[ 
\infer{Q\act{\alpha} Q_1\quad Q\act{\beta} Q_2\quad \alpha \dcl \beta} 
{\xst{Q_3,Q_4}{( \ Q_1 \wact{\beta\minus\alpha} Q_3,\quad  
 Q_2 \wact{\alpha\minus\beta} Q_4, \quad 
 Q_3\wbis Q_4 \ )} } 
\] 
\end{definition} 

It is easy to produce programs
which are locally confluent but not confluent. For instance,
$A = \ol{s}_{1} \isum B$ where $B = \ol{s}_{2} \isum A$.
However, one may notice that this program is {\em not} reactive.
Indeed, for reactive programs local confluence is equivalent
to confluence.

\begin{theorem}\label{det-conf-thm}
\Defitemf{(1)}
A program is determinate if and only if it is confluent.

\Defitem{(2)}
A reactive program is determinate if and only if for all its derivatives $Q$:
\[
\infer{Q\act{\alpha} Q_1, \quad Q\act{\alpha} Q_2,\quad\alpha\in\set{\tau,N}}
{\xst{Q_3,Q_4}{(Q_1 \wact{\tau} Q_3,\quad Q_2\wact{\tau} Q_4,\quad Q_3\wbis Q_4)}}
\]
\end{theorem}

The fact that confluent programs are
determinate is standard and it essentially follows from the
observation that confluent programs are $\tau$-inert.  The observation
that determinate programs are confluent is specific of the
$\spi$-calculus and it depends on the remark that input and output
actions automatically commute with the other compatible
actions.\footnote{We note that the commutation of the inputs 
arises in the $\pi$-calculus with asynchronous communication too, while the
commutation of the outputs is due to the fact that messages on signals
unlike messages on channels persist within an instant (for instance, in
CCS, if $P= \ol{a} \mid a.\ol{b}$ then
$P\act{\ol{a}} a.\ol{b}$, $P\act{\tau} \ol{b}$, and there is no way
to close the diagram).}

The part (2) of the theorem is proved as follows.
First one notices that the stated conditions are equivalent to local
confluence (again relying on the fact that commutation of input and
output actions is automatic) and then following \cite{GS96}
one observes that local confluence plus reactivity entails confluence.

We conclude this section by noticing a strong 
commutation property of $\tau$ actions
that suffices to entail $\tau$-inertness and
determinacy. 
Let $\acteq{\alpha}$ be $\act{\alpha}\union \w{Id}$ where
$\w{Id}$ is the identity relation.

\begin{proposition}\label{strong-confluence}
A  program is determinate if for all its derivatives $Q$:
\[
\infer{Q\act{\tau} Q_1, \quad Q\act{\tau} Q_2}
{\xst{Q'}{(Q_1 \acteq{\tau} Q',\quad Q_2\acteq{\tau} Q')}}
\qquad
\infer{Q\act{N} Q_1, \quad Q\act{N} Q_2}
      {Q_1\wbis Q_2}
\]
\end{proposition}

This is proven by showing that the strong commutation 
of the $\tau$-actions entails $\tau$-inertness.

\section{Conclusion}
We have developed a framework to analyse the determinacy of programs
in a {\em synchronous} $\pi$-calculus.  First, we have introduced a
compositional notion of labelled bisimulation.  Second, we have
characterised a relevant contextual bisimulation as a standard
bisimulation over a modified labelled transition system.  Third, we
have studied the notion of confluence which turns out to be equivalent
to determinacy, and we have shown that under reactivity, confluence
reduces to a simple form of local confluence.

According to theorem \ref{det-conf-thm}(2), there
are basically two situations that need to be analysed in order to
guarantee the determinacy of (reactive) programs.  (1) At least two
distinct values compete to be received within an instant, for
instance, consider: $\emit{s}{v_{1}} \mid \emit{s}{v_{2}} \mid
s(x).P,K$.  (2) At the end of the instant, at least two distinct
values are available on a signal. For instance, consider:
$\emit{s}{v_{1}} \mid \emit{s}{v_{2}} \mid \s{pause}.A(!s)$.  Based on
this analysis, we are currently studying an {\em affine} type system in the
style of \cite{KPT99} that avoids completely the first situation and
allows the second provided the behaviour of the continuation $A$ does
not depend on the order in which the values are collected.

\appendix

\section{Basic properties of labelled bisimulation}\label{basic-prop}
We collect some basic properties of the notion of labelled
bisimulation.  First, we consider a standard variation of the
definition \ref{def-bis} of bisimulation where transitions are
weak on both sides of the bisimulation game.

\begin{definition}[w-bisimulation]
A symmetric relation $\rel{R}$ on programs is a w-bisimulation
if 
\[
\infer{P\rl{R} Q,\quad P\wact{\alpha} P', \quad \w{bn}(\alpha)\inter \w{fn}(Q)=\emptyset}
{\xst{Q'}{( \ Q\wact{\alpha} Q',\quad P'\rl{R} Q' \ )}}
\]
\smallskip\noindent
We denote with $\wbis_{w}$ the largest w-bisimulation.
\end{definition}

With respect to this modified definition we introduce the usual notion
of bisimulation up to bisimulation.\footnote{We recall that it
is important that this notion is defined with respect to w-bisimulation.
Indeed, proposition \ref{basic-lemma}(3) below fails if w-bisimulation is replaced
by bisimulation.}

\begin{definition}[w-bisimulation up to w-bisimulation]
A symmetric relation $\rel{R}$ on programs is a w-bisimulation up to w-bisimulation
if 
\[
\infer{P\rl{R} Q,\quad P\wact{\alpha} P', \quad \w{bn}(\alpha)\inter \w{fn}(Q)=\emptyset}
{\xst{Q'}{( \ Q\wact{\alpha} Q',\quad P'\wbis_{w} \comp \rl{R} \comp \wbis_{w} Q' \ )}}
\]
\smallskip\noindent
We denote with $\wbis_{w}$ the largest w-bisimulation.
\end{definition}

\begin{proposition}\label{basic-lemma}
\Defitemf{(1)} 
The relation $\wbis$ is an equivalence relation.

\Defitem{(2)}
The relations $\wbis$ and $\wbis_{w}$ coincide.

\Defitem{(3)}
If $\rel{R}$ is a w-bisimulation up to w-bisimulation
then $\rel{R}\subseteq \wbis_{w}$.
\end{proposition}
\Proof 
\Proofitemf{(1)} The identity relation is a labelled bisimulation and
the union of symmetric relations is symmetric. To check
transitivity, we prove that $\wbis \comp \wbis$ is 
a labelled bisimulation by standard diagram chasing.

\Proofitem{(2)} 
By definition a w-bisimulation is a labelled bisimulation, therefore
$\wbis_w \subseteq \wbis$. To show the other inclusion, prove
that $\wbis$ is a w-bisimulation again by a standard diagram chasing.

\Proofitem{(3)} First note that by (1) and (2), 
it follows that the relation $\wbis_{w}$ is transitive. 
Then one shows that if $\rel{R}$ is a w-bisimulation up to 
w-bisimulation then the relation 
$\wbis_{w} \comp \rl{R} \comp \wbis_{w}$ is a w-bisimulation. \qed

\subsection{Structural equivalence}\label{struct-equiv}
In the diagram chasing arguments, it will be convenient to 
consider programs up to a notion of 
`structural equivalence'.
This is the least equivalence relation $\equiv$ such that 
(1) $\equiv$ is preserved by static contexts, 
(2) parallel composition is associative and commutative,
(3) $\nu s \ (P\mid Q) \equiv \nu s \ P \mid Q$ if $s\notin \w{fn}(Q)$,
(4) $\emit{s}{v} \mid \emit{s}{v} \equiv \emit{s}{v}$, and 
(5) $\emit{s}{e} \equiv \emit{s}{v}$ if $e\eval v$.
One can check for the different labelled transition systems we consider
that equivalent programs generate exactly the same transitions and
that the programs to which they reduce are again equivalent.

\section{Proof of theorem \ref{compose-bis-thm}}\label{compose-bis-proof}
The theorem follows directly from the following lemma \ref{lemma-wbis-congr}(4).

\begin{lemma}\label{lemma-wbis-congr}

\Defitem{(1)} If $P_1\lbis P_2$ and $\sigma$ is an injective renaming
then $\sigma P_1 \lbis \sigma P_2$.

\Defitem{(2)}
The relation $\lbis$ is reflexive and transitive.

\Defitem{(3)}
If $P_1 \lbis P_2$ then $(P_1 \mid \emit{s}{v}) \lbis (P_2 \mid \emit{s}{v})$.

\Defitem{(4)} If $P_1\lbis P_2$ then $\new{s}{P_{1}} \lbis \new{s}{P_{2}}$ and
$(P_1\mid Q) \lbis (P_2 \mid Q)$.

\end{lemma}
\Proof 
\Proofitemfb{(1), (2)} Standard arguments.

\Proofitemb{(3)} 
Let $\cl{R'}=\set{((P\mid \emit{s}{v}),(Q\mid \emit{s}{v})) \mid
P\wbis Q}$ and $\cl{R} = \cl{R'}\union \wbis$.
We show that $\cl{R}$ is a bisimulation.
Suppose $(P\mid\emit{s}{v})\act{\alpha} \cdot$
and  $P\wbis Q$.
There are two interesting cases to consider.

\Defitem{(\alpha=\tau)} 
Suppose $(P\mid \emit{s}{v}) \act{\tau} (P'\mid \emit{s}{v})$
because $P\act{s?v} P'$.
By definition of the lts, we have
that $P\act{sv} (P\mid \emit{s}{v}) \act{\tau} (P'\mid \emit{s}{v})$.
By definition of bisimulation, 
$Q \wact{sv} (Q''\mid \emit{s}{v}) \wact{\tau} (Q'\mid \emit{s}{v})$ 
and $(P'\mid \emit{s}{v}) \wbis (Q'\mid \emit{s}{v})$.
We conclude, by noticing that then 
$(Q\mid \emit{s}{v}) \wact{\tau} (Q'\mid \emit{s}{v})$.

\Defitem{(\alpha=N)}
Suppose $(P\mid \emit{s}{v})\act{N} P'$. Notice that
$P\act{sv} (P\mid \emit{s}{v})$. Hence:
\[
Q\wact{sv}(Q''\mid \emit{s}{v}) \wact{\tau}
(Q'''\mid \emit{s}{v}) \act{N} Q', \quad
(P\mid \emit{s}{v}) \wbis (Q''\mid \emit{s}{v}) \wbis (Q'''\mid \emit{s}{v}),
\quad \mbox{and} \quad P'\wbis Q'~.
\]
Then $(Q\mid \emit{s}{v}) \wact{N} Q'$.

\Proofitemb{(4)} 
We show that  
$\rel{R}=\set{(\new{\vc{t}}{(P_1 \mid Q)}, \new{\vc{t}}{(P_2\mid Q)})  \mid
P_1 \lbis P_2} \union \lbis$ is a  labelled bisimulation up to 
the structural equivalence $\sbis$.

\Proofitem{(\tau)} Suppose $\new{\vc{t}}{(P_1 \mid Q)} \act{\tau}\cdot$.
This may happen because either $P_1$ or $Q$ perform a $\tau$ action
or because $P_1$ and $Q$ synchronise. We analyse the various
situations.

\Defitem{(\tau)[1]} Suppose $Q\act{\tau} Q'$. Then
$\new{\vc{t}}{(P_2 \mid Q)} \act{\tau}\new{\vc{t}}{(P_2 \mid Q')}$
and we can conclude.

\Defitem{(\tau)[2]} Suppose $P_1 \act{\tau} P'_1$. Then 
$P_2\wact{\tau} P'_2$ and $P'_1 \lbis P'_2$.
So $\new{\vc{t}}{(P_2 \mid Q)} \wact{\tau}\new{\vc{t}}{(P'_2 \mid
  Q)}$  and we can conclude.

\Defitem{(\tau)[3]} Suppose 
$P_1 \act{s?v} P'_1$ and $Q\act{\outact{t'}{s}{v}}Q'$.
This means $Q\equiv \nu \vc{t'} \ (\emit{s}{v} \mid Q'')$ and
$Q'\equiv (\emit{s}{v} \mid Q'')$. By (3), 
$(P_1 \mid \emit{s}{v}) \wbis (P_2 \mid \emit{s}{v})$.
Moreover, $(P_1 \mid \emit{s}{v}) \act{\tau} (P'_1\mid\emit{s}{v})$.
Therefore, $(P_2\mid \emit{s}{v}) \wact{\tau} (P'_2 \mid \emit{s}{v})$ and
$(P'_1\mid\emit{s}{v}) \wbis  (P'_2 \mid \emit{s}{v})$.
Then we notice that the transition
 $\new{\vc{t}}{(P_1 \mid Q)} \act{\tau}\cdot \equiv 
\new{\vc{t},\vc{t'}}{((P'_1 \mid \emit{s}{v}) \mid Q'')}$ 
is matched by the transition 
 $\new{\vc{t}}{(P_2 \mid Q)} \act{\tau}\cdot \equiv 
\new{\vc{t},\vc{t'}}{((P'_2 \mid \emit{s}{v}) \mid Q'')}$.

\Defitem{(\tau)[4]} Suppose $P_1 \act{\outact{t'}{s}{v}} P'_1$ and 
$Q\act{s?v} Q'$. 
Then  $P_2 \wact{\outact{t'}{s}{v}} P'_2$ and $P'_1 \wbis P'_2$. 
And we conclude noticing that 
$\new{\vc{t}}{(P_2 \mid Q)}\wact{\tau} \new{\vc{t},\vc{t'}}{(P'_2 \mid
  Q')}$. 

\Proofitem{(\w{out})} Suppose 
$\new{\vc{t}}{(P_1 \mid Q)} \act{\outact{t'}{s}{v}} \cdot$.
Also assume $\vc{t}=\vc{t_{1}},\vc{t_{2}}$ and
$\vc{t'} = \vc{t_{1}},\vc{t_{3}}$ up to reordering so that
the emission extrudes exactly the names $\vc{t}_{1}$ among the names in 
$\vc{t}$. We have two subcases depending which component performs the action.

\Defitem{(\w{out})[1]} Suppose 
 $Q \act{\outact{t_{3}}{s}{v}} Q'$. 
Then $\new{\vc{t}}{(P_2 \mid Q)} \act{\outact{t'}{s}{v}} 
\new{\vc{t_{2}}}{(P_2 \mid Q')}$ and we can conclude.

\Defitem{(\w{out})[2]} Suppose $P_1 \act{\outact{t_{3}}{s}{v}} P'_1$.
Then $P_2 \wact{\outact{t_{3}}{s}{v}} P'_2$ and 
$P'_1 \lbis P'_2$. Hence 
$\new{\vc{t}}{(P_2 \mid Q)} \wact{\outact{t'}{s}{v}} 
\new{\vc{t_{2}}}{(P'_2 \mid Q)}$ and we can conclude.

\Proofitem{(\w{in})}  It is enough to notice
that, modulo renaming,
$\new{\vc{t}}{(P_i \mid Q)} \mid \emit{s}{v} \equiv
\new{\vc{t}}{((P_i\mid \emit{s}{v}) \mid Q)}$ and
recall that by (3), $(P_1\mid \emit{s}{v}) \wbis (P_2\mid \emit{s}{v})$.

\Proofitem{(N)} Suppose  $\new{\vc{t}}{(P_1 \mid Q)} \susp$. 
Up to structural equivalence, we can express $Q$ as
$\new{\vc{t}_Q}{(S_Q \mid I_Q)}$ where $S_Q$ is the
parallel composition of emissions and $I_Q$ is the
parallel composition of receptions.
Thus we have:
$\new{\vc{t}}{(P_1 \mid Q)}  \sbis
 \new{\vc{t},\vc{t}_Q}{(P_1 \mid S_Q \mid I_Q)}$,
and 
$\new{\vc{t}}{(P_2 \mid Q)} \sbis
 \new{\vc{t},\vc{t}_Q}{(P_2 \mid S_Q \mid I_Q)}$
assuming $\set{\vc{t}_{Q}}\inter \w{fn}(P_i)=\emptyset$
for $i=1,2$.

If $\new{\vc{t}}{(P_1 \mid Q)} \act{N} P$ then
$P\sbis \new{\vc{t},\vc{t}_Q}{(P''_1 \mid Q')}$ where in
particular, we have that
$(P_1\mid S_Q)\dcl$ and $(P_1 \mid S_Q) \act{N}
(P'_1 \mid 0)$.

By the hypothesis $P_1\lbis P_2$,  and by definition of bisimulation 
we derive that: 
(i) $(P_2 \mid S_Q) \wact{\tau} (P''_{2} \mid S_Q)$,
(ii) $(P''_{2} \mid S_Q)\susp$,
(iii) $(P''_{2} \mid S_Q)\act{N} (P'_{2}  \mid 0)$,
(iv) $(P_1\mid S_Q)\lbis (P''_{2} \mid S_Q)$, and
(v) $(P'_1 \mid 0) \lbis (P'_{2}  \mid 0)$.

Because $(P_1\mid S_Q)$ and $(P''_{2} \mid S_Q)$ are
suspended and bisimilar, the two programs must
commit (cf. definition \ref{commit-def}) on the same signal names
and moreover on each signal name they must emit the same set of
values up to renaming of bound names. It follows that the program
$\new{\vc{t},\vc{t}_Q}{(P''_{2} \mid S_Q \mid I_Q)}$ is
suspended. The only possibility for an internal transition 
is that an emission in $P''_{2}$ enables a reception in $I_Q$ but
this contradicts the hypothesis that 
$\new{\vc{t},\vc{t}_Q}{(P_{1} \mid S_Q \mid I_Q)}$ 
is suspended.
Moreover, $(P''_{2} \mid S_Q \mid I_Q)\act{N}
(P'_{2} \mid 0 \mid Q' )$.

Therefore, we have that 
\[
\new{\vc{t}}{(P_2 \mid Q)} \sbis
 \new{\vc{t},\vc{t}_Q}{(P_2 \mid S_Q \mid I_Q )}
 \wact{\tau}
 \new{\vc{t},\vc{t}_Q}{(P''_2 \mid S_Q \mid I_Q)},
\]
$\new{\vc{t},\vc{t}_Q}{(P''_2 \mid S_Q \mid I_Q)}\susp$, 
and 
$\new{\vc{t},\vc{t}_Q}{(P''_2 \mid S_Q \mid I_Q)} \act{N}
 \new{\vc{t},\vc{t}_Q}{(P'_2 \mid 0 \mid Q')}$.
Now  $\new{\vc{t},\vc{t}_Q}{(P_1 \mid S_Q \mid I_Q)}
      \rl{R}
      \new{\vc{t},\vc{t}_Q}{(P''_2 \mid S_Q \mid I_Q)}$
because $(P_1 \mid S_Q)\lbis (P''_2 \mid S_Q)$ and
$\new{\vc{t},\vc{t}_Q}{(P'_1 \mid Q')}
      \rl{R}
      \new{\vc{t},\vc{t}_Q}{(P'_2 \mid Q')}$
because $P'_1 \lbis P'_2$. \qed

\section{Proof of theorem \ref{char-bis}}\label{char-bis-proof}
We start with the labelled transition system defined in table \ref{Lts} and
the notion of bisimulation in definition \ref{def-bis}.
In table \ref{equiv-bis}, we incrementally modify the labelled
transition system and/or the conditions in the bisimulation game.
This leads to three equivalent characterisations of the notion of
bisimulation. We prove this fact step by step.

\begin{lemma}\label{char-lemma-1}
The bisimulation $\wbis$ coincides with the bisimulation $\wbis_1$.
\end{lemma}
\Proof
The only difference here is in the rule $(\w{in}_{\w{aux}})$, the
bisimulation conditions being the same. Now this rule produces an
action $s?v$ and the latter is an auxiliary action that is used to
produce the relevant action $\tau$ thanks to the rule $(\w{synch})$.
A simple instance of the difference follows. Suppose
$P=\emit{s}{e} \mid s(x).Q,K$ and $e\eval v$. Then:
\[
\begin{array}{l}
P \act{\tau} \emit{s}{e} \mid [v/x]Q=P' \mand
P \actI{\tau} \emit{s}{e} \mid ([v/x]Q \mid \emit{s}{v}) = P''~.
\end{array}
\] 
In the $\spi$-calculus, we do not distinguish 
the situations where the same value is emitted once or more times
within the same instant. In particular, $P'$ and $P''$ 
are structurally equivalent (cf. section \ref{struct-equiv}). \qed 
\\

\noindent
Next, we focus on the relationships between the labelled transitions
systems $\actI{\w{act}}$ and $\actII{\w{act}}$. 
In $\actII{\w{act}}$, the rule $(\w{in})$ is removed and in the
rule $(\w{in}_{\w{aux}})$, the label $s?v$ is replaced by
the label $sv$ (hence the auxiliary action $s?v$ is not used in 
this labelled transition system).

\begin{lemma}\label{lts12}
\Defitemf{(1)}
If $P\actI{\w{act}} P'$ and $\w{act}\neq sv$ 
then $P\actII{\w{act'}} P'$ 
where $\w{act'}=sv$ if $\w{act}=s?v$, 
and $\w{act'}=\w{act}$ otherwise.

\Defitem{(2)}
If $P\actII{\w{act}} P'$  
then $P\actI{\w{act'}} P'$ 
where $\w{act'}=s?v$  if $\w{act}=sv$,
and $\w{act'}=\w{act}$ otherwise.
\end{lemma}

We also notice that 1-bisimulation is preserved by parallel
composition with an emission; the proof is similar
to the one of lemma \ref{lemma-wbis-congr}(3).

\begin{lemma}\label{emission-cong}
If $P\wbis_1 Q$ then 
$(P\mid \emit{s}{v}) \wbis_1 (Q\mid \emit{s}{v})$.
\end{lemma}

\begin{lemma}\label{char-lemma-2}
The bisimulation $\wbis_1$ coincides with the bisimulation $\wbis_2$.
\end{lemma}
\Proof
\Proofitemf{(\wbis_1 \subseteq \wbis_2)}
We check that $\wbis_1$ is a 2-bisimulation.
If $\alpha=sv$ then we apply lemma \ref{emission-cong}.
Otherwise, suppose $\alpha\neq sv$, $P\wbis_1 Q$,  and
$P\actII{\alpha} P'$.
By lemma \ref{lts12}(2), $P\actI{\alpha} P'$.
By definition of 1-bisimulation, $\xst{Q'}{Q\wactI{\alpha} Q',
  P'\wbis_1 Q'}$.
By lemma \ref{lts12}(1), $Q\wactII{\alpha}Q'$.

\Proofitem{(\wbis_2 \subseteq \wbis_1)}
We check that $\wbis_2$ is a 1-bisimulation.
If $\alpha=sv$ and $P\actI{sv}(P\mid \emit{s}{v})$ then by definition
of the lts, $Q\actI{sv} (Q\mid \emit{s}{v})$. 
Moreover, by definition of 2-bisimulation, $(P\mid \emit{s}{v})
\wbis_2 (Q\mid \emit{s}{v})$.
Otherwise, suppose $\alpha\neq sv$, $P\wbis_2 Q$,  and
$P\actI{\alpha} P'$.
By lemma \ref{lts12}(1), $P\actII{\alpha} P'$.
By definition of 2-bisimulation, $\xst{Q'}{Q\wactII{\alpha} Q',
  P'\wbis_2 Q'}$.
By lemma \ref{lts12}(2), $Q\wactI{\alpha}Q'$. \qed \\

Next we move to a comparison of 2 and 3 bisimulations.
Note that both definitions share the same lts denoted with
$\actII{\alpha}$.  First we remark the following.

\begin{lemma}\label{lemma23-prelim}
\Defitem{(1)} If $P\wbis_2 Q$ and $P\act{N} P'$ then
$\xst{Q', Q''}{( \ Q\wactII{\tau} Q'', Q'' \act{N} Q',
P\wbis_2 Q'', P'\wbis_2 Q' \ )}$.

\Defitem{(2)} If $P\wbis_3 Q$ then $(P\mid \emit{s}{v}) \wbis_3 
(Q\mid \emit{s}{v})$.
\end{lemma}
\Proof \Proofitemf{(1)}
If $P\act{N} P'$ then $P$ cannot perform $\tau$ moves.
Thus if $P\wbis_2 Q$ and $Q\wactII{\tau} Q''$ then
necessarily $P \wbis_2 Q''$.

\Proofitem{(2)} Again we follow the proof of lemma \ref{lemma-wbis-congr}(3).
Let $\cl{R'}=\set{((P\mid \emit{s}{v}),(Q\mid \emit{s}{v})) \mid
P\wbis_3 Q}$ and $\cl{R} = \cl{R'}\union \wbis_3$.
We show that $\cl{R}$ is a 3-bisimulation.
Suppose $(P\mid\emit{s}{v})\actI{\alpha} \cdot$ 
and  $P\wbis_3 Q$.
There are two interesting cases to consider.

\Proofitem{(\alpha=\tau)} 
Suppose $(P\mid \emit{s}{v}) \actII{\tau} (P'\mid \emit{s}{v})$
because $P\actII{sv} P'$.
By definition of 3-bisimulation, 
either (i) $Q\wactII{sv} Q'$ and $P'\wbis_3 Q'$ or 
(ii) $Q\wactII{\tau} Q'$ and $P'\wbis_3 (Q'\mid  \emit{s}{v})$.
In case (i), $(Q\mid \emit{s}{v}) \wact{\tau} (Q'\mid \emit{s}{v})$ 
and we notice that $((P'\mid \emit{s}{v}),(Q'\mid \emit{s}{v}))\in
\cl{R}$.
In case (ii), $(Q\mid \emit{s}{v}) \wact{\tau} (Q'\mid \emit{s}{v})$ 
and we notice that $(P'\mid \emit{s}{v},(Q'\mid \emit{s}{v}) \mid
\emit{s}{v})\in \cl{R}$ and $(Q'\mid \emit{s}{v}) \mid
\emit{s}{v} \equiv (Q'\mid \emit{s}{v})$.

\Proofitem{(\alpha =N)}
Suppose $((P\mid \emit{s}{v}) \mid S) \act{N} P'$.
By definition of 3-bisimulation, taking $S'= (\emit{s}{v} \mid S)$
$(Q\mid S') \wact{\tau} Q''\act{N} Q'$,
$(P\mid S') \wbis_3 Q''$, and  $P'\wbis_3 Q'$.  \qed

\begin{lemma}\label{bis23-lemma}
The bisimulation $\wbis_2$ coincides with the bisimulation $\wbis_3$.
\end{lemma}
\Proof
\Proofitemf{(\wbis_2 \subseteq \wbis_3)}
We show that $\wbis_2$ is a 3-bisimulation. We look first at the
condition for the input.
Suppose $P\wbis_2 Q$ and $P\actII{sv} P'$.
By definition of 2-bisimulation, $(P\mid \emit{s}{v}) \wbis_2 (Q\mid \emit{s}{v})$.
Also $(P\mid \emit{s}{v}) \actII{\tau} (P'\mid \emit{s}{v}) \equiv P'$.
By definition of 2-bisimulation, $(Q\mid \emit{s}{v}) \wact{\tau}
(Q'\mid \emit{s}{v})$ and $P'\equiv (P'\mid \emit{s}{v}) \wbis_2
(Q'\mid \emit{s}{v})$. Two cases may arise.

\Defitem{(1)} If $Q\wact{sv} Q'$ then $Q'\mid \emit{s}{v} \equiv Q'$ and 
we satisfy the first case of the input condition for 3-bisimulation.

\Defitem{(2)} If $Q\wact{\tau} Q'$ then, up to structural equivalence,
we satisfy the second case of the input condition for 3-bisimulation.
\smallskip  

\noindent
Next we consider the condition for the end of the instant.
Suppose  $P\wbis_2 Q$,  $S=\emit{{s}_{1}}{v_{1}} \mid \cdots \mid
\emit{{s}_{n}}{v_{n}}$, and $(P\mid S) \actII{N} P'$.
By condition $\w{(Inp)}$, $(P\mid S) \wbis_2 (Q\mid S)$.
Then, by lemma \ref{lemma23-prelim}(1), the condition of
3-bisimulation  is entailed by the corresponding condition for
2-bisimulation applied to $(P\mid S)$ and $(Q\mid S)$.

\Proofitem{(\wbis_3 \subseteq \wbis_2)}
We show that $\wbis_3$ is a 2-bisimulation. 
The condition $(\w{Inp})$ holds because of lemma
\ref{lemma23-prelim}(2).
The condition of 2-bisimulation for the end of the instant is a
special case of the condition for 3-bisimulation where we take
$S$ empty. \qed

\section{Proof of theorem \ref{det-conf-thm} and proposition \ref{strong-confluence}}\label{det-conf-proof}
First, relying on proposition \ref{basic-lemma}(3), one can repeat
the proof in \cite{Milner89} that confluence implies $\tau$-inertness 
and determinacy.

\begin{proposition}\label{conf-det} 
If a program is confluent then it is $\tau$-inert and determinate. 
\end{proposition} 
\Proof 
Let  $\cl{S}=\set{(P,P') \mid P \mbox{ confluent and }P\wact{\tau} P' }$ and 
define $\cl{R}=\cl{S} \union \cl{S}^{-1}$.
We show that $\cl{R}$ is a w-bisimulation up to
w-bisimulation (cf. lemma \ref{basic-lemma}(3)). Clearly $\cl{R}$ is symmetric.
Then suppose $P$ confluent and $P\wact{\tau} Q$ 
(the case where $Q$ reduces to $P$ is symmetric).  
If $Q\wact{\alpha} Q_1$ then $P\wact{\alpha} Q_1$ and $Q_1 \rl{R} Q_1$.
On the other hand, if $P\wact{\alpha} P_1$
then by confluence there are $P_2,Q_1$ such  
that $P_1\wact{\tau} P_2$, $Q \wact{\alpha} Q_1$, 
and $P_2\wbis Q_1$. Thus $P_1 \rl{R} \comp \wbis Q_1$.
 
Therefore if $P$ is confluent and $P\wact{\tau} P'$ then $P\wbis P'$.  Also
recall that if $Q$ is a derivative of $P$ then $Q$ is confluent.  Thus
we can conclude that if $P$ is confluent then it is $\tau$-inert.

Next, we show that: 
\[ 
\infer{P_1 \wbis P_2, \quad  
P_1 \wact{\alpha} P_3,  \quad
P_2 \wact{\alpha} P_4} 
{P_3\wbis P_4}~.  
\] 
By definition of bisimulation, $\xst{P_5}{( \ P_2 \wact{\alpha} P_5, 
P_3 \wbis P_5 \ )}$. By confluence, 
$\xst{P_6,P_7}{( \ P_5\wact{\tau} P_6, P_4\wact{\tau} P_7, P_6\wbis 
  P_7 \ )}$. 
By $\tau$-inertness and transitivity, $P_3 \wbis P_4$. 

Finally, we can iterate this observation to conclude that if 
$P\wact{\alpha_{1}} \cdots \wact{\alpha_{n}} P_1$ and
$P\wact{\alpha_{1}} \cdots \wact{\alpha_{n}} P_2$ then
$P_1\wbis P_2$. \qed  \\

We pause to point-out the particular properties of the input and
output actions in the labelled transition system in table \ref{Lts}.  
It is easily verified that if $P\act{\nu \vc{t} \ol{s}{v}} P'$ then
$P\equiv \nu \vc{t}(\emit{s}{v} \mid P'')$ and $P' \equiv (\emit{s}{v}
\mid P'')$.  This entails that in the following lemma the cases that
involve an output action are actually general up to structural
equivalence. 

\begin{lemma}[input-output commutations]\label{io-comm-lemma}~
{\small
\[
\begin{array}{lc}

(\w{in}-\tau)

&
\infer{P\act{sv} (P\mid \emit{s}{v}), \quad P\act{\tau} P'}
{(P\mid \emit{s}{v})\act{\tau} (P'\mid \emit{s}{v}), \quad 
  P' \act{sv} (P'\mid \emit{s}{v})} \\ \\

(\w{in}-\w{in})

&
\infer{P\act{sv} (P\mid \emit{s}{v}), \quad 
P\act{s'v'} (P\mid \emit{s'}{v'})}
{\begin{array}{c}
(P\mid \emit{s}{v})\act{s'v'} (P\mid \emit{s}{v})\mid
\emit{s'}{v'},\quad
(P\mid \emit{s'}{v'})\act{sv} (P\mid \emit{s'}{v'})\mid
\emit{s}{v}, \\
(P\mid \emit{s}{v})\mid
\emit{s'}{v'} \equiv (P\mid \emit{s'}{v'})\mid \emit{s}{v}
\end{array}} \\ \\

(\w{out}-\tau)

&\infer{\nu\vc{t} (\emit{s}{v} \mid P) \act{\outact{t}{s}{v}}
  (\emit{s}{v}\mid P), \quad
\nu\vc{t} (\emit{s}{v} \mid P) \act{\tau} \nu\vc{t} (\emit{s}{v} \mid
P')}
{(\emit{s}{v}\mid P) \act{\tau} (\emit{s}{v} \mid P'), \quad 
\nu\vc{t} (\emit{s}{v} \mid P') \act{\outact{t}{s}{v}}  (\emit{s}{v} \mid P')}
\\ \\ 

(\w{out}-\w{in})

&\infer{\nu\vc{t} (\emit{s}{v} \mid P) \act{\outact{t}{s}{v}}
(\emit{s}{v}\mid P),\quad 
\nu\vc{t} (\emit{s}{v} \mid P) \act{s'v'} \nu\vc{t} (\emit{s}{v} \mid
P) \mid \emit{s'}{v'}}
{(\emit{s}{v}\mid P) \act{s'v'} (\emit{s}{v}\mid P) \mid \emit{s'}{v'},\quad
\nu\vc{t} (\emit{s}{v} \mid P) \mid \emit{s'}{v'} 
\act{\outact{t}{s}{v}} (\emit{s}{v} \mid
P) \mid \emit{s'}{v'}}
\\ \\

(\w{out}-\w{out})

&
\infer{
\begin{array}{c}
\nu\vc{t} (\emit{s_{1}}{v_{1}} \mid \emit{s_{2}}{v_{2}} \mid P) 
\act{\outact{t_{1}}{s_{1}}{v_{1}}}
\nu \vc{t}\minus \vc{t_{1}} \ (\emit{s_{1}}{v_{1}} \mid
\emit{s_{2}}{v_{2}} \mid P),\\
\nu\vc{t} (\emit{s_{1}}{v_{1}} \mid \emit{s_{2}}{v_{2}} \mid P) 
\act{\outact{t_{2}}{s_{2}}{v_{2}}}
\nu \vc{t}\minus \vc{t_{2}} \ (\emit{s_{1}}{v_{1}} \mid
\emit{s_{2}}{v_{2}} \mid P)
\end{array}
}
{
\begin{array}{c}
\nu \vc{t}\minus \vc{t_{1}} \ (\emit{s_{1}}{v_{1}} \mid
\emit{s_{2}}{v_{2}} \mid P)
\act{\outact{t_{2} \minus t_{1}}{s_{2}}{v_{2}}}
(\emit{s_{1}}{v_{1}} \mid
\emit{s_{2}}{v_{2}} \mid P)
,\\
\nu \vc{t}\minus \vc{t_{2}} \ (\emit{s_{1}}{v_{1}} \mid
\emit{s_{2}}{v_{2}} \mid P)
\act{\outact{t_{1} \minus t_{2}}{s_{2}}{v_{2}}}
(\emit{s_{1}}{v_{1}} \mid
\emit{s_{2}}{v_{2}} \mid P)
\end{array}}

\end{array}
\]}
\end{lemma}

Note that, up to symmetry (and structural equivalence), 
the previous lemma covers  {\em all} possible commutations
of two compatible actions $\alpha,\beta$ 
but the 2 remaining cases where $\alpha=\beta$ and 
$\alpha\in \set{\tau,N}$.

\begin{proposition}\label{prop-det-conf}
If a program is  deterministic then it is confluent.
\end{proposition}
\Proof
We recall that if $P$ is deterministic then it is $\tau$-inert.
Suppose $Q$ is a derivative of $P$, $\alpha \dcl \beta$, 
$Q\wact{\alpha} Q_1$ and $Q\wact{\beta} Q_2$. 

If $\alpha=\beta$ then the definition of determinacy implies that
$Q_1\wbis Q_2$. Also note that $\alpha\minus \beta =\beta\minus\alpha
=\tau$ and $Q_i\wact{\tau} Q_i$ for $i=1,2$. So the conditions for
confluence are fulfilled.

So we may assume $\alpha\neq \beta$ and, up to symmetry, we are left
with 5 cases corresponding to the 5 situations considered in lemma
\ref{io-comm-lemma}. 

In the 2 cases where $\beta=\tau$ we have that $Q\wbis Q_2$ by
$\tau$-inertness.
Thus, by bisimulation $Q_2 \wact{\alpha} Q_3$ and $Q_1\wbis Q_3$.
Now $\alpha\minus \tau =\alpha$, $\tau\minus \alpha =\tau$, and
$Q_1 \wact{\tau} Q_1$. Hence the conditions for confluence are
fulfilled.

We are left with 3 cases where $\alpha$ and $\beta$ are distinct
input or output actions. By using $\tau$-inertness, we can focus
on the case where $Q\wact{\alpha} Q_1$ and $Q\act{\beta} Q'_2\wact{\tau} Q_2$.
Now, by iterating the lemma \ref{io-comm-lemma}, 
we can prove that:
\[
\infer{Q\ (\act{\tau})^n \ Q'_1,\quad n\geq 1, \quad Q\act{\beta} Q'_2}
{\xst{Q''_2}{(\ Q'_1 \act{\beta} Q''_2,\quad Q'_2 \ (\act{\tau})^n \ Q''_2 \ )}}~.
\]
So we are actually reduced to consider the situation where
$Q\act{\alpha} Q' _1\wact{\tau} Q_1$ and $Q\act{\beta}Q'_2 \wact{\tau} Q_2$.
But then by lemma \ref{io-comm-lemma}, we have:
$Q'_1 \act{\beta\minus\alpha} Q_3$,
$Q'_2 \act{\alpha\minus \beta} Q_4$, and $Q_3\equiv Q_4$.
Then using $\tau$-inertness and bisimulation, it is easy to close the
diagram. \qed \\

This concludes the proof of the first part of the theorem 
(\ref{det-conf-thm}(1)). 
To derive the second part, we rely on
the following fact due to \cite{GS96}.
 
\begin{fact}[\cite{GS96}]\label{gs-thm}
If a program is reactive and locally confluent then it is confluent. 
\end{fact}

Thus to derive the second part of the theorem (\ref{det-conf-thm}(2))
it is enough to prove.

\begin{proposition}
A program is locally confluent if (and only if) 
for all its derivatives $Q$:
\[
\infer{Q\act{\alpha} Q_1, \quad Q\act{\alpha} Q_2,\quad\alpha\in\set{\tau,N}}
{Q_1 \wact{\tau} Q_3\quad Q_2\wact{\tau} Q_4\quad Q_3\wbis Q_4}
\]
\end{proposition}
\Proof 
The stated condition is a special case of local confluence thus
it is a necessary condition.
To show that it is sufficient to entail local confluence, it is enough 
to appeal again to lemma \ref{io-comm-lemma} 
(same argument given at the end of 
the proof of proposition  \ref{prop-det-conf}). \qed

\paragraph{Proof of proposition \ref{strong-confluence}}
Say that $P$ is {\em strong confluent} if it satisfies the hypotheses
of proposition \ref{strong-confluence}.  Let $\cl{S}= \set{(P,Q) \mid
P \mbox{ strong confluent and } (P\equiv Q \mbox{ or } P\act{\tau}
Q)}$. Let $\cl{R}= \cl{S} \union \cl{S}^{-1}$.  We show that $\cl{R}$
is a bisimulation.  Hence strong confluence entails $\tau$-inertness.
Note that if $P\act{\alpha}P_i$, for $i=1,2$, and $\alpha$ is either
an input or an output action then $P_1\equiv P_2$.
By lemma \ref{io-comm-lemma} and diagram chasing, we show that if $P$
is strong confluent and $P\wact{\alpha}P_i$, for $i=1,2$, then
$P_1\wbis P_2$.  
This suffices to show that $P$ is determinate (and
confluent). \qed

\end{document}